# KULR

## Kashmir University Law Review

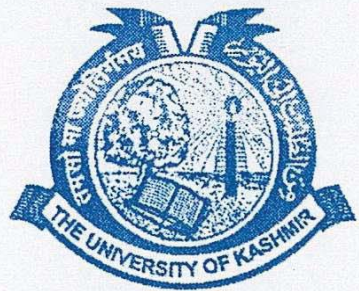

## Reprint

| VOL. XIII | 2006 | No. XIII |

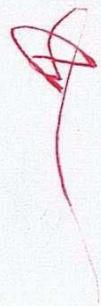

**Department of Law**

**University of Kashmir Srinagar**

# SPAM – Technological and Legal Aspects

*M. Tariq Banday* [*] *& Jameel A. Qadri* [**]

**Abstract**

In this paper an attempt is made to review technological, economical and legal aspects of the spam in detail. The technical details will include different techniques of spam control e.g., filtering techniques, Genetic Algorithm, Memory Based Classifier, Support Vector Machine Method, etc. The economic aspect includes Shaping/Rate Throttling Approach/Economic Filtering and Pricing/Payment based spam control. Finally, the paper discusses the legal provisions for the control of spam. The scope of the legal options is limited to USA, European Union, New Zealand, Canada, Britain and Australia.

**Introduction**

The opening of Electronic Commerce by the US congress in 1995 lead to a significant increase in the use of Internet by individuals and companies with email as the most used application. The flexibility provided by the Open Medium on the Internet for communication and sharing of information made it possible to sent unsolicited bulk email (UBE) and unsolicited commercial email (UCE) also referred to as spam. There is no single agreed definition of spam, and in fact one challenge in debating the issue of spam is defining it. Some believe it as any commercial email send to a recipient without the prior affirmative consent of the recipient. To others, it is commercial email to which affirmative or implied consent was not given. Implied consent may mean a pre-existing business relationship. For others spam is an unwanted commercial email. Whether or not a particular email is unwanted, of course, varies per recipient. Since senders of UCE do find buyers for some of their products, it can be argued that at least some UCE is reaching interested consumers, and therefore is wanted, and thus is not spam. Consequently, some argue that marketers should be able to send commercial email messages as long as they allow each recipient an opportunity to indicate that in future such emails are not desired. Another group considers spam to be only fraudulent commercial email, and believe that commercial email messages from legitimate senders should only be permitted.

---

[*] Lecturer, Department of Electronics and Instrumentation Technology, The University of Kashmir, Hazratbal, Srinagar, India, email: sgrmtb@yahoo.com.

[**] School of Computing, Middlesex University, Hendon, London, UK, email: scorpiojameel@yahoo.com.

*Spam and its variants*

Loosely defining spam is any message or posting, regardless of its content, that is sent to multiple recipients who have not specifically requested the message (Opt-In). Spam can also be multiple postings of the same message to newsgroups or list servers that are not related to the topic of discussion. A person engaged in spamming is called spammer. Spam in blogs called blog spam or comment spam is a form of search engine spamming done manually or automatically by posting random comments, promoting commercial services, to blogs, wikis, guestbooks, or other publicly-accessible online discussion boards. Any web application that accepts and displays hyperlinks submitted by visitors may be a target of Link Spam. This is the placing or solicitation of links randomly on other sites, placing a desired keyword into the hyperlinked text of the backlink. Blogs, guest books, forums and any site that accepts visitors' comments are particular targets and are often victims of drive-by spamming, where automated software creates nonsense posts with links that are usually irrelevant and unwanted. Link spam dishonestly and deliberately manipulates link-based ranking algorithms of search engines like Google's PageRank to increase the rank of a web site or page so that it is placed as close to the top of search results as possible. Phishing is a particular type of spam which reflects social engineering. Phishing frauds are characterized by attempts to masquerade as a trustworthy person or emulate an established, reputed business in an electronic communication such as email. The objective is to trick recipients into divulging sensitive information such as bank account numbers, passwords, and credit card details. A person engaged in phishing activities is called a phisher. Spam generally refers to email, rather than other forms of electronic communication. The term spim, for example, is used for unsolicited advertising via Instant Messaging. Spit refers to unsolicited advertising via Voice Over Internet Protocol (VOIP). Unsolicited advertising on wireless devices such as cell phones is called wireless spam.

*Motivation behind sending spam*

Spam arises from an online social situation that has been created by technology. First, it costs little more to send a million email messages than to send one. Second, hits are a percentage of transmissions, so sending more spam means more sender profit. In 2002, The Wall Street Journal published a profile of a successful spammer who resided in a 5,000 square feet house in Florida and had a projected annual income of $200,000. Successful spammers only need to receive 100 responses out of 10 million messages sent to earn a profit. The actual cost for a veteran spammer to produce a successful spamming campaign can be as little as $50 for every one million recipients, a total cost of only $.00005 per email sent. Spam generators thus logically seek to reach all users. Without responders, spam would never exist, but a small fraction of recipients

always respond. Spammers need only 100 takers per 10 million requests to earn a profit[1] which means a percentage much lesser than 0.01 percent hit rate. Even if spam blockers successfully blocks 99.99% of all spam, spammer transmissions would still continue to increase.

*Tools used for spamming*

Spammers use various techniques and tools that include spoofing, Bot-Networks, Open Proxies, Open mail relays, Untraceable Internet connections, and Bulk Email tools for sending unsolicited bulk email[2]. Spammers operate as a creative group who work through secret networks to meet and share email addresses. The two most effective approaches to gather or harvest emails are to monitor Internet use and to ask for email addresses. Some of the most popular methods of email harvesting are Guessing, Purchased Lists, Legitimate Email Lists, Web Pages, White/Yellow Page Sites, Web & Paper Forms, UseNet Posts, Web Browser, Hacking, User Profiles, IRC and Chat Rooms.

*Ruinous results of spam*

It is difficult to put spam problem in actual figures as the opinions vary significantly. At one extreme, authors fear that spam will overwhelm users and make them stop using email altogether. By the year 2015 volume of spam will exceed 95% of all email traffic[3], and people will give up email accounts in frustration while a few like Crews[4] argue that spam email is a minor annoyance. The volume of unwanted email is lowering the productivity of email users by an estimated 1.4%-3.1%[5].The cost of transporting and delivering spam is not borne by spammers, but by ISP's and ultimately passed on to the end users[6]. There are indirect costs as well. Email is now the vehicle for delivering viruses and phishing attacks, which can lead to data loss, financial losses or even identity theft. Also, the use of filtering to reduce spam has led to the risk of false positives where legitimate and sometimes very important emails do not get delivered. Email senders use its low cost and easy access to send information to recipients who have not requested it. This disclosure may increase the recipient's welfare if the information is valuable or decrease it if the data is fraudulent or useless. Email reverses the normal cost pattern of communication as sending is cheap relative to receiving. Email creates low costs for a sender. An advertiser bears hardware cost, Internet access fee, acquisition of recipient addresses, composition of message contents, mechanism to capture revenue, and risk. Most of these costs are fixed and others are generally low. Thus, sending additional spam incurs trivial marginal costs. Even if marginal revenue is low, spammers have an incentive to send more messages to cover fixed initial investments. Thus, an advertiser who transmits spam tends to send a lot of it. Recipients bear the majority of email's costs because most recipients receive email through an ISP (Internet Service Provider) and ISPs act as spam aggregation points as they receive spam for many users from many senders. Thus, ISPs pass the costs of processing this large email volume on to recipients and the architecture of Internet email generates a cost-multiplier effect for aggregators, such as ISPs, due to recipient processing. Recipients typically bear hardware cost, message storage costs, Internet access fee, message processing or filtering cost, time to read and delete messages, risk of missing desired messages due to volume of unwanted ones, risk factor, potential harm from fraudulent messages such as phishing, etc., Psychological effects, such as from viewing pornographic spam or from the annoyance of managing large volumes of unwanted mail. Recipients, unlike senders, face non-trivial marginal costs for additional messages. ISPs in particular must devote considerable storage space, processing, and reputation investments to deal with spam. Users must download and then delete unwanted messages, and may open messages containing offensive material. As the spam volume increases, the risk of missing non-spam messages in one's inbox increases.

---

1 A. Weiss, "Ending Spam's Free Ride," netWorker, vol. 7, no. 2, 2003, pp. 18-24.

2 L. McLaughlin. Bot software spreads, causes new worries. IEEE Distributed Systems Online, (6):1541–4922, June 2004. http://csdl.computer.org/comp/mags/ds/2004/06/o6001.pdf.

3 B. Whitworth and E. Whitworth, "Spam and the social technical gap," IEEE Computer, vol. 37, no. 10, pp. 38-45, Oct. 2004.

4 C.W. Crews, Jr., "Why canning "spam" is a bad idea," Cato Policy Analysis, no. 408, July 26, 2001; http://www.cato.org/pubs/pas/pa408.pdf.

5 Spam: The serial ROI killer, Res. note E50, June 2004; http://www.nucleusresearch.com/ research/e50.pdf.

6 P. Dougan, "Legal and technical responses to unsolicited commercial email ('spam')," 2001; http://www.strath.ac.uk/Other/staffclub/web2law/spam.pdf

Given to offensive, fraudulent, and adult-oriented material that offends recipients and that parents want to protect their children from seeing, the extensive resources consumed, high costs of processing and other misleading content of spam, the unanimous opinion suggests that spam is a grave problem and a collective solution that includes technological, legal and social measures be undertaken to put a check on it.

## Technological Aspects

Email is one of the earliest Internet applications; originally it contemplated users sending messages directly to each other's screens in addition to each other's mailboxes. Internet pioneer Jon Postel formalized the technical specifications for transferring email with the Simple Mail

Transfer Protocol (SMTP) which has undergone several revisions later in 1981[7] and thereafter. The Internet Engineering Task Force (IETF) adopted SMTP as one of its Requests for Comments (de facto standards for Internet applications) in RFC 821. SMTP is an application layer protocol for TCP/IP based Internet infrastructure. An Internet email exchange that conforms to SMTP's RFC 821 is often compared to a conversation between two parties, the sender and the receiver described by RFC 821 as a "lock-step" exchange. The sender has information an email message; it wants to transfer to the receiver. SMTP defines how the sender and receiver communicate to transfer this information in essence; the protocol creates a set of conversational and grammatical rules for exchanging email between computers. The SMTP exchange is highly structured and ordered; when the sender makes a statement, the receiver must respond before the sender can make another statement. After each of the sender's statements, the receiver either accepts the information or rejects it with an error code that indicates the problem. The SMTP conversation involves creation of connection, establishment of the SMTP conversion, defining the message sender, defining the message recipient(s) and transferring the message content. This system uses servers and clients (software and hardware subsystems) on sender and receiver sides along the path of an email message. Four interrelated components: sender client, sender server, receiving server, and receiving client are used. Outlook Express and Eudora running on a client computer are examples of email clients while Exchange and Sendmail running on a server computer are examples of email servers. In web based email services such as in case of mail.yahoo.com and gmail.google.com clients and servers are combined and are integrated behind a web server. An email sender uses their client side to compose a message. The client connects to the sending server and delivers the message in an outgoing queue. Next, the sending server connects to the receiving server, validates the existence of the recipient on the receiving server, and delivers the message. The message is stored on the receiving server until retrieved by its addressee. Eventually, the human recipient uses the client on their side to connect to the receiving server and retrieve the message.

The SMTP is simple in content and requirements. It minimizes information that must be included in the exchange and leaves functions such as authentication to other protocols and applications. This simple architecture makes SMTP easy to implement and use, but the spammers have exploited these advantages and have used them for sending spam email messages[8]. Spammer takes advantage of the trust built into RFC 821. SMTP defines how computers communicate to send and receive email, but it does not define when not to use it. When two computers use SMTP to transfer email, their default behavior is to accept each other's representations. Spammers exploit this underlying trust to target recipients, hide their own identities, and conceal their tracks. Email system uses a set of open standards; this reliance constitutes both a great strength and an inherent weakness of the medium. Email standards are defined and maintained by the IETF (Internet Engineering Task Force), a non-profit organization dedicated to creating universally accessible protocols for Internet uses. Open standards reduce coordination costs for vendors and assure interoperability. A software company that wants to create a new email server or client does not need to obtain a license for the core technologies involved. Adoption of change in an open standard based system can not be compelled. Thus far, email software vendors have not sought to fix the spam problem within SMTP; rather, their solutions treat the protocol as given. The IETF offers protocols that add security features to SMTP, but these have not been widely adopted. Anti-spam proposals such as Caller-ID for Email and Sender Policy Framework (SPF) work through the DNS rather than changing SMTP. The backwards-compatibility challenge and the need for widespread, if not universal, adoption of any solution, impede the effort to revise SMTP to help solve the spam problem. Different entities on the Internet have proposed various types of technological solutions to mitigate the damage caused by spam which are presented next.

---

7 Klensin, J., "Simple Mail Transfer Protocol," RFC 2821, April 2001.

8 Hoffman, SMTP Service Extension for Secure SMTP over Transport Layer Security (Feb. 2002), at http://www.ietf.org/rfc/rfc3207.txt?number=3207.

*Server and Client Based Solutions*

Many researchers are presently working in the implementation of new filters that prevent spam from reaching their destination either by blocking it at the server level or at the client level. Any technique that can reduce the volume of spam at the receiving client will reduce costs associated with productivity loss for the human recipient. However, all costs associated with delivering the message, storing it on the receiving server, and delivering it to the client are borne by the owner of the receiving server, and ultimately passed on to the end user. These costs can be reduced if spam is stopped before it reaches the receiving server. The ideal case is when spam never even leaves the sending client. The spam control techniques operate at various points in the transmission link. The closer to the sending client these techniques stop spam, the more efficient they would be. Server based solutions are expected to control the injection, propagation, and delivery of spam, and in turn reduce the traffic load incurred by spam. Server based solutions focus on the configuration of mail servers to prevent unauthorized usage as default blind-relay nodes[9]. IP based port filtering [10] can be used to block SMTP ports for incoming/outgoing

---

9 Mogul, J. C. 1989. Simple and flexible datagram access controls for UNIX-based gateways. In USENIX Conference Proceedings. Baltimore, MD, 203–221.

messages from/to unauthorized servers. Also, ISPs are expected to handle complaints from spam victims and take appropriate action on spammer accounts. Unfortunately, ISPs are normally reluctant to strengthen such policies. Protecting email users individually can have its advantages; however, a server-based solution is normally more advantageous, especially for organizations. For email users who have a spam problem with their personal email account, using a client-based filter can reduce the number of spam messages received, therefore, can appear to solve their problem. However, an organization who wishes to reduce its spam should explore other options. In most cases, dealing with the problem at a server level is a better choice[11]. Filtering that is done on the receiver client is easiest to deploy by an end user, but least effective, because spam has already used up network and storage resources. Filtering on the receiving server can drop spam even before it is stored locally. Client-based solutions do not prevent an organization's networks from being taxed by unwanted emails[12]. Whether at the client level or at the server level, there are several techniques for filtering the spam. Technical anti-spam approaches comprise of one or more of the following basic approaches:

*Munging*

Munging is to deliberately alternate an email address to make it unusable for email harvesters, who build email lists for spamming purposes. For example, tariq@yahoo.com could be munged as tariq at yahoo dot com. Intuitively speaking, munging only provides a weak defense line in preventing email addresses from being harvested. It could temporarily fool most of the web-based spambots, which are programs designed to collect email addresses from Internet in order to build mailing lists to send spam mails. However, it is not hard for spammers to adapt all sorts of munging tricks.

*Origin or Authentication Based Filtering Techniques*

These techniques are also called **Access Filtering Techniques.** Origin-based filters use network information in order to detect spam. IP and email address are the most common pieces of network information used. The four major types of origin-based filters are blacklists, whitelists, greylist and challenge/response systems. Some protocol based spam filtering techniques discussed later in this paper are also based on origin or authentication.

*Blacklist*

---

10 Lindberg, G. 1999. Anti-spam recommendations for SMTP MTAs. RFC 2505, Network WorkingGroup. Feb.

11 TrimMail Inbox, http://www.trimmail.com/.

12 J.A. Korsmeyer, Inc, www.jak.com.

Blacklists, also known as realtime blackhole lists (RBL) or domain name system black lists (DNSBL), can filter mail from mail servers or domains that have sent spam or are suspected of doing so. IP addresses of known or suspected spammers are entered into centrally maintained databases and made available as blacklists through the Internet. Standard DNS lookups are used to query these databases at the time of SMTP connection or when mail is received, with spam classification occurring based on the reply given. Blacklists are managed by various separate groups; each with its own focus and different policies in regards to how an IP address gets on (and off) the list[13]. As blacklists only require DNS lookups, they have a very low CPU overhead and are generally easy to implement. Another advantage of blacklists is that they allow spam to be blocked at the SMTP connection phase, effectively stopping it from entering the network. These lists could potentially be removed at any time without warning, leaving networks solely relying on these blacklists without any form of spam protection at all. The effectiveness of a blacklist relies on the people who manage them; if blacklists are not updated in a timely manner, spam can get through. Spammers can circumvent blacklists to a certain degree by using zombie networks. As a zombie network comprises of many different computers, all of which could be from different domains, a blacklist on a specific domain would provide only minimal spam protection.

*Whitelist*

Whitelists allow users to specifically define "trusted" addresses that will immediately classify as legitimate any email received from those addresses. An appealing quality of whitelists is that for most users a whitelist would be significantly smaller and easier to maintain than a blacklist. Also, mail flagged by a whitelist as legitimate can bypass further through spam filters, effectively reducing the load on those filters. Since the sender of email messages is not authenticated, spammers who can guess an address on the whitelist can then freely propagate spam to that address. Whitelists are, therefore, best used when combined with other spam blocking techniques[14].

*Greylist*

---

13 Allman, E 2003, 'Spam, Spam, Spam, Spam, Spam, the FTC, and Spam', Queue, vol. 1, no. 6, pp. 62-9.

14 Garcia, FD, Hoepman, J-H & van Nieuwenhuizen, J 2004, 'Spam Filter Analysis', paper presented to 19th IFIP International Information Security Conference, Toulouse, France.

Greylisting temporarily reject mails from unfamiliar senders (e.g. not in a whitelist) and require the rejected mails to be retransmitted[15]. A properly configured Mail Transfer Agent, as suggested in IETF RFC 2821, should retransmit mails in 30 minutes after a failure. However, spam mails sent through an open proxy or a non-properly-configured MTA will not be retransmitted. As an example, Matador from Mail-Frontier holds incoming emails in a greylist until senders respond with correct answers on certain questions.

*Challenge/Response Systems*

Challenge/response systems are an advanced version of whitelists, allowing senders who are not on the whitelist to have their emails received. Incoming messages from addresses not on the whitelist trigger an automatic reply (or challenge) to the sender, requiring them to prove that they are a real user and not an automated mailer. For example, the sender may be required to click on a link in the reply message and enter a valid email address and the ID number of the response message. If this process is completed, then the email successfully passes through the challenge/response system[16]. The challenge/response method aims to protect against automated mailer programs by forcing the user to complete a task that is simple for a human but too complicated for a program to handle. Challenge/response systems also protect against spammers who manually send email, as the time required to complete the challenge could be better used sending spam to additional addresses. Challenge/response systems also help to protect against the generally large amount of false positives generated by traditional whitelist systems. One problem with whitelists is the issue of deadlock. If two parties who have never corresponded before both run challenge/response systems, the challenge sent by the recipient's system will be caught by the sender's challenge/response system and neither party will have the opportunity to provide an appropriate response. This problem could be alleviated if the original sender adds the recipient's address to their whitelist before commencing communication. Another problem associated with the use of challenge/response systems is legitimate automated email lists that the user has subscribed to. These lists cannot respond to the challenge messages generated by the system, and mail from these sources may be marked as spam. As with the deadlock issue, this problem could be alleviated if the subscriber adds the mailing list address to their whitelist before subscribing to the list.

*Content Filtering*

---

15 Evan Harris, The Next Step in the Spam Control War: Greylisting, Aug. 21, 2003, available at http://projects.puremagic.com/greylisting/whitepaper.html.

16 Pfleeger, SL & Bloom, G 2005, 'Canning Spam: Proposed Solutions to Unwanted Email', Security & Privacy Magazine, IEEE, vol. 3, no. 2, pp. 40-7.

Content filtering detects spam by looking inside the email and examining the message context. Most content based spam detection systems try to understand the text to various extends in order to identify spam. Content filtering includes rank based or heuristic filters that search the email message for pattern that indicate spam. These patterns could include specific words or phrases, malformed message headers and large amount of exclamation marks and capital letters. The problem with the rule based spam filtering method is that, since the rule set is largely static, they are easily defeatable by spammers' technique such as word obfuscation. Statistical detection techniques assign a point value or a spam probability rating calculated with a formula. Emails are separated into tokens with each token assigned a probability value. These probabilities are then combined to find the email spam probability. The interesting aspect of the statistical spam detection methods is their continuous ability to learn and thus are also known as machine learning methods which include Naïve Bayes (NB), term frequency – inverse document frequency (TF-IDF), K-nearest neighbor (K-NN), and support vector machines (SVMs).

*Ṇaïve Bayes Method*

The Naïve Bayes (NB) classifier is a probability-based approach. The basic concept of it is to find whether an email is spam or not by looking at which words are found in the message and which words are absent from it. In the literature, the NB classifier for spam is defined as follows:

$$C_{ṆB} = \arg\max_{C_i \varepsilon T} P(C_i) \prod_k P(w_k | C_i)$$

Where $T$ is the set of target classes (spam or non-spam), and $P(w_k|C_i)$ is the probability that word $w_k$ occurs in the email, given the email belongs to class $C_i$. The likelihood term is estimated as

$$P(w_k | C_i) = \frac{n_k}{Ṇ}$$

Where $n_k$ is the number of times word $w_k$ occurs in emails with class $C_i$, and $Ṇ$ is the number of words in emails with class $C_i$.

To calculate an email's spam probability with a good degree of accuracy, Bayesian filters need to be "trained" by being given examples of what constitutes a spam email and what does not. The advantage of this technique is that, given appropriate time and training data, Bayesian filters can achieve a combination of extremely high accuracy rates with a low percentage of false positives[17]. A further advantage of Bayesian filters is that they are constantly self-adapting and if provided ongoing training data from the user, Bayesian filters evolve to stop new spam techniques.

---

17 Graham, P 2003, 'Better Bayesian Filtering', paper presented to 2003 Spam Conference.

*Term Frequency-Inverse Document Frequency Method*

The most often adopted representation of a set of messages is term weight vectors which are used in the Vector Space Model[18]. The term weights are real numbers indicating the significance of terms in identifying a document. Based on this concept, the weight of a term in an email message can be computed by the *tf . idf* . The *tf* (term frequency) indicates the number of times that a term *t* appears in an email. The *idf* (inverse document frequency) is the inverse of document frequency in the set of emails that contain t.

The *tf . idf* weighting scheme is defined as

$$w_{ij} = tf_{ij} \cdot \log(\frac{I}{df_i})$$

Where $w_{ij}$ is the weight of the $i^{th}$ term in the $j^{th}$ email, $tf_{ij}$ is the number of times that the $i^{th}$ term occurs in the $j^{th}$ email, $I$ is the total number of emails in the collection, and $df_i$ is the number of emails in which the $i^{th}$ term occurs.

*K-nearest neighbor method*

The most basic instance-based method is the K-nearest neighbor (K-NN) algorithm. It is a very simple method to classify documents and to show very good performance on text categorization tasks. If we want to apply K-NN method to classify emails, the emails of the training set have to be indexed and then convert them into a document vector representation. When classifying a new email, the similarity between its document vector and each one in the training set has to be computed. Then, the categories of the k nearest neighbors are determined and the category which occurs most frequently is chosen.

*Support Vector Machine Method*

Support vector machine (SVM) has become very popular in the machine learning community because of its good generalization performance and its ability to handle high-dimensional data by using kernels. An email may be represented by a feature vector **x** that is composed of the various words from a dictionary formed by analyzing the collected emails. Thus, an email is classified as spam or non-spam by performing a simple dot product between the features of an email and the SVM model weight vector, $y = \mathbf{w} \cdot \mathbf{x} - b$, where y is the result of classification, **w** is weight vector corresponding to those in the feature vector **x**, and b is the bias parameter in the SVM model that is determined by the training process.

*Memory Based Classifier*

---

18 G. Salton, Automating Text Processing: The Transformation, Analysis and Retrieval of Information by Computer, Addison-Wesley, 1989.

Memory-based classifier, which was suggested to combine multiple ground-level classifiers, a.k.a. stacked generalization to induce a higher-level classifier for improving overall performance in antispam filtering. The solution is a hierarchical approach where the high-level classifier can be considered as the president of a committee with the ground-level classifiers as members.

*Genetic Algorithm*

It is an adaptive spam mail filtering technique which uses genetic algorithm and its operations, i.e., crossover and mutation, to create new varieties of spam mail prototypes. The experiments show that proposed adaptive spam mail filtering performs efficient results. In additional, the system allows the threshold value for matching rules to be set manually for appropriate level of filtering.

*Shaping/Rate Throttling Approach/Economic Filtering*

The main attractiveness of spamming is that sending large amounts of small email messages is relative cheap compared to other marketing techniques, the idea behind shaping filters is to make sending high volumes of email traffic to be more expensive, thereby making it less attractive to send bulk emails. This forces the junk email sender to spend considerable computing resources to send a spam messages. More effective in reducing overall spam volume are a host of techniques that aim to reduce spam before it can reach the receiving server. Teergrubing[19] is the process of delaying the receipt of a message. As the sender server contacts a teergrubing receiving server to deliver a message, the receiving server delays answering requests. Other similar ideas include TarProxy[20] and Jackpot. Diffmail[21] is another shaping approach to delay the receipt of mail from unknown senders.

*Pricing/Payment based spam control*

Another class of spam control techniques is based on increasing the costs to spammers, proportional to the volume of email sent[22]. A sender is required to post a financial bond, certifying that they will not spam. Any recipient that believes she has received spam from that sender can request that the sender be debited for a compensation amount. The service is free to email receivers, and costs senders an application fee, an annual licensing fee and a per spam complaint fee that is deducted from the amount of the posted bond and transferred to "an independent, disinterested nonprofit organization"[23]. Instead of this fixed payment, Dai and Li[24] propose a dynamic pricing mechanism that would control the flow of spam to maximize the recipient's utility function. Yet another similar approach is zmail[25] that requires no definition of what is spam and what is not spam, so spammers' efforts to evade such definition become irrelevant.

---

19 L. Donnerhacke, Teergrubing FAQ: http://www.iks-jena.de/mitarb/lutz/usenet/teergrube.en.html.

20 Open Source Technology Group, 2006; http://sourceforge.net/projects/tarproxy.

21 A Differential Message Delivery Architecture to Control Spam, Zhenhi Duan, Yingfei Dong, Kartik Gopalan, 11th International Confrence on Parallel and Distributed Systems, IEEE 2005.

22 Kraut, R. E., Sunder, S., Telang, R., and Morris, J. (2005). Pricing electronic mail to solve the problem of spam. Human-Computer Interaction, 20, nnn-nnn.

*Information Hiding/Identity Hoping/Aliasing Based Spam Control*

A possible solution to spam is to hide one's email address, by using one-time email addresses[26]. Users would generate a temporary alias for the purpose of a particular transaction, and would retire the address if spam starts to arrive via that address. Several companies offer this service (Spamex.com, Sneakemail.com and Spammotel.com are among them). Ad-hoc email address is used in this method. Tsuyoshi Abe[27] has proposed an ad-hoc email address service system which employs a cryptographic algorithm for the generation of ad-hoc addresses. The addresses are associated with mail filter rules and subscribers' original address undisclosed to public. This algorithm enables mail transfer agents to filter incoming mails by extracting filter rules from ad-hoc addresses, and thus requires no database to store filter rules for thread addresses.

*Protocol Based Spam Control Techniques*

Protocol based solutions include the use of digital signatures for authentication. Some spam prevention techniques that broadly fall into the category of origin-based filters are sender authentication systems such as the Sender ID Framework[28] and DomainKeys[29]. These systems place more emphasis on the authentication of the sender or the sender's domain but require modifications to the existing email system to be effective thus are classified in protocol based spam control techniques. **Digital signatures,** also known as fingerprints, identify messages. Signatures of messages that have been identified as spam can be put in a database. This database is then used to compare the signature of received email with the list of signatures of spam. If there is a match, the email is spam. Ideally, this would mean that once one person that uses this database receives a particular spam message, the signature is added and that message would be blocked for all the other users of the database. Unfortunately, it does take time for a signature to be added to the database; therefore, others will receive the email before it is blocked. Furthermore, slightly modifying a message will change its signature. Spammers often add random text in their emails which changes their digital signature[30]. The purpose is that people who can be authenticated can also be held accountable for their email practices. Several joint projects between major Internet companies operate in this arena, including Sender Policy Framework[31]. Domain Keys Identified Mail (DKIM) supported by Cisco and Yahoo, and Microsoft's Sender ID. These solutions operate at the receiving server and give higher priority to email from authenticated senders. The receiving server checks whether the sender is a valid server in the domain where it claims to be. This check is based on looking up information in the Domain Name Server (DNS) databases distributed on the Internet (the technique is generally referred to as reverse DNS lookup or reverse MX record lookup, RMX). The direct feedback to the sender server refused or delayed email if the sender IP address is not authenticated – makes this type of technique very effective in reducing spam.

---

23 Return Path, Inc., 2006; www.bonded-sender.com.

24 R. Dai and K. Li "Shall we stop all unsolicited email messages?" in Proc. First Conference on Email and Anti-Spam (CEAS 2004), Mountain View, CA, 2004; http://www.ceas. cc/papers-2004/189.pdf.

25 Zmail: zero-sum Free market Control of Spam, Benjman J. Kuipers, Alex X. Liu, Ashin Gautam, Mohmed G. Gouda, 25th IEEE conference on Distributed Computing Systems Workshop, 2005.

26 J.-M. Seigneur and C.D. Jensen, "Privacy recovery with disposable email addresses," IEEE Security & Privacy Mag., vol. 1, no. 6, pp. 35-39, 2003.

27 Spam Filtering with cryptographic Ad-hsoc email Addresses, Tsuyoshi Abe, Jun Miyake, Masahisa Kawashima, Katsumi Takahashi, IEEE 2005, (SAINT-W'2005).

---

28 Lyon, J & Wong, M 2004, Work in Progress, Internet-Draft: Sender ID: Authenticating Email, Internet Engineering Task Force, <http://download.microsoft.com/download/6/c/5/6c530 77f-013e-480c-a19d-78785 0d84861/senderid_spec1.pdf>.

29 Delany, M 2005, Work in Progress, Internet-Draft: Domain-based Email Authentication Using Public-Keys Advertised in the DNS (DomainKeys), Internet Engineering Task Force, <http://www.ietf.org/internet-drafts/draft-delanydomainkeys-base-02.txt>.

30 Paul Graham, Better Bayesian filtering, January 2003, http://paulgraham.com/better.html.

31 M. W. Wong, "Important considerations for implementers of SPF and/or Sender ID;" http://www.maawg.org/about/whitepapers/spfsendID/.

*Collaborative Spam Control Techniques*

Most anti-spam solutions currently in use involve a combination of the above listed techniques. Several vendors sell anti-spam appliances that combine multiple techniques, sometimes in a collaborative enterprise level or distributed fashion, and that allow remote management. The performance of collaborative technique tends to be superior to that of individual techniques. Collaborative spam filters use the collective memory of, and feedback from, users to reliably identify spam. Some collaborative spam filters, such as SpamNet (www.cloudmark.com), deliver performance comparable to that of Bayesian filters. A major drawback of collaborative filtering schemes is that they ignore the already present and pervasive social communities in cyberspace and instead try to create new ones of their own to facilitate information sharing. The most well known collaborative filtering systems are Vipul's Razor (razor.sourceforge.net) and Distributed Checksum Clearinghouse (www.rhyolite.com/anti-spam/dcc/). Spammers have adapted to such collaborative techniques by making slight changes in messages from one recipient to another, but collaborative techniques still work by identifying commonalities across messages.

*Other Approaches*

The above listed filtering techniques is an exhaustive list but not a complete one, other data mining, machine learning, text classification are under research which include digest-based filters, density-based filters, chi-squared filters, global collaboration filters and artificial neural networks. Social network techniques, such as Reputation Network Analysis and other types namely Ontology Driven Spam Filters, Rough Set Theory Based Filters, Image Analysis based filtering are also under research.

## Legal Solutions

Policing of virtual world through law has experienced several problems. First, current law assumes physical world architecture. But virtual world works differently, and physical laws may not transfer easily. Second, virtual worlds frequently change faster than legislators can draft laws. Spam has already shown it can mutate into new forms, like spim. Each spam variant would require new laws. Yet society takes years to pass them, while Internet applications can change in months. Third, in cyberspace code is law, so the programmers who write the code make the rules. Giving spammers' anonymity, or the power to hide their source, they can negate any law. Finally, jurisdiction limits laws, as attempts to legislate telemarketers illustrate.

Although several countries have framed anti-spam laws, yet the CAN-SPAM Act of the USA is being referred the most for several reasons that include; USA is considered to be on of the most SPAM producing countries, was one of the first to introduce laws against unsolicited commercial email and most of the spammers operate from USA.

*Federal Law (CAṆ -SPAM Act)*

In USA nine bills were introduced in the 108[th] Congress for regulating spam, of which most were "opt-out" bills. The Congress passed S.877 (Burns-Wyden) Bill with merged provisions from several House and Senate Bills[32]. Several amendments were later made to this Bill and finally it was enacted into the law Controlling the Assault on Non-Solicited Pornography and Marketing (CAN-SPAM) Act P.L. 108-187. The Law CAN-SPAM Act became effective in US from 1[st] January; 2004.The CAN-SPAM Act is an opt-out law, requiring senders of all commercial emails to provide a legitimate opt-out opportunity to recipients. Several anti-spam groups were of the opinion that commercial email should not be send to the receivers unless they opt-in. It is worth mentioning here that European Union had adopted an opt-in requirement for email that became effective on 31[st] October, 2003. However, not all countries of European Union use Opt-in requirement. Several US states passed Spam Laws. The CAN-SPAM Act preempts state spam laws, but not other state laws that are not specific to electronic mail, such as contract, tort law, trespass or other state laws to the extent that they relate to fraud or computer crime. California passes spam law that would have become effective on January 1, 2004 and was considered relatively strict. It required opt-in for UCE unless there was a prior business relationship, in which case, opt-out is required. The anticipated implementation of that California law is often cited as one of the factors that stimulated Congress to complete action on a less restrictive, preemptive federal law before the end of 2003.

*CAṆ -SPAM Act: An Analysis*

- Commercial email may be sent to recipients as long as the message conforms to the following requirements
    a)   Transmission information in the header is not false or misleading,
    b)   Subject headings are not deceptive,
    c)   (A functional Opt-Out System). A functioning return email address or comparable mechanism is included with the Commercial email message to

---

32 Nine bills were introduced in the 108th Congress prior to passage of the CAN-SPAM Act: H.R. 1933 (Lofgren), H.R. 2214 (Burr-Tauzin-Sensenbrenner), H.R. 2515 (Wilson-Green), S. 877 (Burns-Wyden), S. 1052 (Nelson-FL), and S. 1327 (Corzine) were "opt-out" bills. S. 563 (Dayton) was a "do not e-mail" bill. S. 1231 (Schumer) combined elements of both approaches. S. 1293 (Hatch) created criminal penalties for fraudulent email.

       enable recipients to indicate they do not wish to receive future commercial email messages from that sender at the email address where the message was received.

   d) The email is not sent to a recipient by the sender, or anyone acting on behalf of the sender, more than 10 days after the recipient has opted-out, unless the recipient later gives affirmative consent to receive the email (i.e., opts back in).,

   e) The email must be clearly and conspicuously identified as an advertisement or solicitation. The legislation however; does not state how or where that identification must be made.

- Commercial email is defined as email, the primary purpose of which is the commercial advertisement or promotion of a commercial product or service (including content on an Internet website operated for a commercial purpose). It does not include transactional or relationship messages. The Act directs the Federal Trade Commission (FTC) to issue regulations within 12 months of enactment of the Act, to define the criteria to facilitate determination of an email's primary purpose.

The FTC issued its final rule on December 16, 2004, exactly one year after the law was enacted. According to the FTC's press release, the final rule clarifies that the Commission does not intend to regulate non-commercial speech. It differentiates between commercial content and "transactional or relationship" content in defining the primary purpose of an email message.

a) If an email contains only a commercial advertisement or promotion of a commercial product or service, its primary purpose is deemed to be commercial.

b) If an email contains both commercial content and transactional or relationship content, the primary purpose is deemed to be commercial if the recipient would likely conclude that it was commercial through reasonable interpretation of the subject line, or if the transactional and relationship content does not appear in whole or in substantial part at the beginning of the body of the message.

c) If an email contains both commercial content, and content that is neither commercial content nor transactional or relationship content, the primary purpose is deemed to be commercial if the recipient would likely conclude that it was commercial through reasonable interpretation of the subject line, or if the recipient would likely conclude the primary purpose was commercial through reasonable interpretation of the body of the message.

d) If an email contains only transactional or relationship content, it is not deemed to be a commercial email message.

"Commercial" content is defined in the final rule as "the commercial advertisement or promotion of a commercial product or service," which includes "content on an Internet website operated for a commercial purpose." That is the same as the definition in the CAN-SPAM Act. (The FTC's notice of proposed rulemaking had a slightly different definition. The final rule emphasizes that, in the final rule, the definition is the same as in the act.) The FTC specifically declined to define the term "spam" because the act sets forth a regulatory scheme built around the terms "commercial electronic mail message" and "transactional or relationship message."

- Some requirements (including the prohibition on deceptive subject headings, and the opt-out requirement) do not apply if the message is a "transactional or relationship message," which include various types of notifications, such as periodic notifications of account balance or other information regarding a subscription, membership, account, loan or comparable ongoing commercial relationship involving the ongoing purchase or use by the recipient of products or services offered by the sender; providing information directly related to an employment relationship or related benefit plan in which the recipient is currently involved, participating, or enrolled; or delivering goods or services, including product updates or upgrades, that the recipient is entitled to receive under the terms of a transaction that the recipient has previously agreed to enter into with the sender. The act allows, but does not require, the FTC to modify that definition.
- Sexually oriented commercial email must include, in the subject heading, a "warning label" to be prescribed by the FTC (in consultation with the Attorney General), indicating its nature. The warning label does not have to be in the subject line, however, if the message that is initially viewable by the recipient does not contain the sexually oriented material, but only a link to it. In that case, the warning label, and the identifier, opt-out, and physical address required under section 5 (a)(5) of the act; must be contained in the initially viewable email message as well. Sexually oriented material is defined as any material that depicts sexually explicit conduct, unless the depiction constitutes a small and insignificant part of the whole, the remainder of which is not primarily devoted to sexual matters. These provisions do not apply, however, if the recipient has given prior affirmative consent to receiving such emails.
- Businesses may not knowingly promote themselves with email that has false or misleading transmission information.
- State laws specifically related to spam are preempted, but not other state laws that are not specific to electronic mail, such as trespass, contract, or tort law, or other state laws to the extent they relate to fraud or computer crime.

- Violators may be sued by FTC, state attorneys general, and ISPs (but not by individuals).
- Violators of many of the provisions of the act are subject to statutory damages of up to $250 per email, to a maximum of up to $2 million, which may be tripled by the court (to $6 million) for "aggravated violations."
- Violators may be fined, or sentenced to up to 3 or five years in prison (depending on the offense), or both, for accessing someone else's computer without authorization and using it to send multiple commercial email messages; sending multiple commercial email messages with the intent to deceive or mislead recipients or ISPs as to the origin of such messages; materially falsifying header information in multiple commercial email messages; registering for five or more email accounts or online user accounts, or two or more domain names, using information that materially falsifies the identity of the actual registrant, and sending multiple commercial email messages from any combination of such accounts or domain names; or falsely representing oneself to be the registrant or legitimate successor in interest to the registrant of five of more Internet Protocol addresses, and sending multiple commercial email messages from such addresses. "Multiple" means more than 100 email messages during a 24-hour period, more than 1,000 during a 30-day period, or more than 10,000 during a one-year period. Sentencing enhancements are provided for certain acts.
- The Federal Communications Commission, in consultation with the FTC, must prescribe rules to protect users of wireless devices from unwanted commercial messages. The act required the FCC to issue regulations concerning spam on wireless devices such as cell phones. The FCC issued those regulations in August 2004.

**The Act does not:**
- Create a "Do Not Email registry" where consumers can place their email addresses in a centralized database to indicate they do not want commercial email. The law requires only that the FTC develop a plan and timetable for establishing such a registry and to inform Congress of any concerns it has with regard to establishing it. The FTC issued its report to Congress on June 15, 2004. The report concluded that without a technical system to authenticate the origin of email messages, a Do Not Email registry would not reduce the amount of spam, and, in fact, might increase it.
- Require that consumers "opt-in" before receiving commercial email.
- Require commercial email to include an identifier such as "ADV" in the subject line to indicate it is an advertisement. The law does require the FTC to report to Congress within 18 months of enactment on a plan for requiring commercial email to be identifiable from its subject line through use of "ADV" or a comparable identifier, or compliance with Internet Engineering Task Force standards, or an explanation of any concerns FTC has about such a plan.

- Include a "bounty hunter" provision to financially reward persons who identify a violator and supply information leading to the collection of a civil penalty, although the FTC must submit a report to Congress within nine months of enactment setting forth a system for doing so. The study was released on September 15, 2004. The FTC concluded that the benefits of such a system are unclear because, for example, without large rewards (in the $100,000 to $250,000 range) and a certain level of assurance that they would receive the reward, whistleblowers might not be willing to assume the risks of providing such information. The FTC offered five recommendations if Congress wants to pursue such an approach. They include:

    a) Tie eligibility for a reward to imposition of a final court order, instead of to collecting a civil penalty.
    b) Fund the rewards through congressional appropriations, instead of through collected civil penalties.
    c) Restrict reward eligibility to insiders with high-value information.
    d) Exempt FTC decisions on eligibility for rewards from judicial or administrative review, and
    e) Establish reward amounts high enough to attract insiders with high value information.

Under the law, the FTC was required to provide Congress with an assessment of the act's effectiveness, and recommend any necessary changes. The FTC submitted its report in December 2005. The FTC concluded that the act has been effective in terms of adoption of commercial email "best practices" that are followed by "legitimate" online marketers, and in terms of providing law enforcement agencies and ISPs with an additional tool to use against spammers. Additionally FTC concluded that the volume of spam has begun to stabilize, and the amount reaching individuals' inboxes has decreased because of improved anti-spam technologies.49 However, it also found that the international dimension of spam has not changed significantly, and that there has been a shift toward the inclusion of "increasingly malicious" content in spam messages, such as "malware," which is intended to harm the recipient. Other negative changes noted by the FTC are that spammers are using increasingly complex multi-layered business arrangements to frustrate law enforcement, and are hiding their identities by providing false information to domain registrars (the "Whois" database). The FTC did not recommend any changes to the CAN-SPAM Act, but encouraged Congress to pass the US SAFE WEB Act (S. 1608, noted that continued consumer education efforts are needed, and called for improved anti-spam technologies, particularly the domain-level authentication. The Undertaking Spam, Spyware, and Fraud Enforcement with Enforcers beyond Borders (U.S. SAFE WEB) Act, was referred to the Senate Commerce Committee. The Senate Commerce Committee ordered reported S. 1608 on December 15, 2005, without amendment.

The US Can-Spam Act is the most famous anti-spam legislation, but not the only one. Following is a brief detail of Spam or the related laws as are enforced in European Union, United Kingdom, Australia, New Zealand and Canada[33].

The *European Union* issued a formal "Privacy and Electronic Communications" Directive[34] in 2002, specifically covering the sending of unsolicited commercial email. Member states are obliged to implement directives in their own national legislation. Indeed, EU directives carry no practical implications on their own, but only when actually transcribed into local laws. Don't expect all EU countries to share common email marketing laws, though, despite the standard established in the EU document. The directive leaves it up to each country to choose their preferred approach on some issues. For example, B2B emails require a mandatory opt-in in Austria, but may be sent to some companies on an opt-out basis in the UK. Countries also differ in how they interpret the wording of the directive. So although the EU documents give you an overview of the base standard in Europe, one still needs to check each country's local email marketing laws on a case-by-case basis. Article 13(1) of the Privacy and Electronic Communications Directive requires Member States to prohibit the sending of unsolicited commercial communications by fax or email or other electronic messaging systems such as SMS and MMS unless the prior consent of the addressee has been obtained (opt-in system). The only exception to this rule is in cases where contact details for sending email or SMS messages (but not faxes) have been obtained in the context of a sale. Within such an existing customer relationship the company who obtained the data may use them for the marketing of similar products or services as those it has already sold to the customer. Nevertheless, even then the company has to make clear from the first time of collecting the data, that they may be used for direct marketing and should offer the right to object. Moreover, each subsequent marketing message should include an easy way for the customer to stop further messages (opt-out). The opt-in system is mandatory for any email, SMS or fax addressed to natural persons for direct marketing. It is optional with regard to legal persons. For the latter category Member States may choose between an opt-in or an opt-out system. For all categories of addressees, legal and natural persons, Article 13(4) of the Directive prohibits direct marketing messages by email or SMS which conceal or disguise the identity of the sender and which do not include a valid address to which recipients can send a request to cease such messages. For voice telephony marketing calls, other than by automated machines, Member States may also choose between an opt-in or an opt-out approach.

---

33 The hyperlink is available at http://www.spamlaws.com/.

34 Data protection law can be viewed on European Union web site at http://ec.europa.eu/justice_home/fsj/privacy/index_en.htm.

The *British* government implemented the relevant EU directive in December 2003 with the Privacy and Electronic Communications Regulations[35]. The legislation has attracted criticism for being too weak, for example by making it legal to send unsolicited email to businesses on a purely opt-out basis. Tougher legislation can be expected in the future. Mainstream advertisers also need to comply with industry self-regulation in the form of various codes of practice.

*Australia's* anti-spam law is the Spam Act 2003[36]. A 2006 review confirmed the usefulness of the legislation such that significant changes are not expected. The Department of Communications, Information Technology and the Arts has conducted a legislative review of the Spam Act. The review is required by legislation to assess the operation of the Spam Act after two years of its operation. In December 2005 the Minister for Communications, Information Technology and the Arts, Senator Helen Coonan, released an issues paper inviting public comment on the operation of the Spam Act. The period for comment closed on 1 February 2006. Sixty four submissions were received, four of which were confidential, from a wide range of industry, consumer and government organisations as well as members of the public. The Department prepared a report based on the submissions received. The Minister tabled the report in Parliament on 22 June 2006.

Legislation regulating commercial email finally passed through the *New Zealand* Parliament at the end of February 2007[37]. The Unsolicited Electronic Messages Bill was introduced in mid-2005 and got its first reading in December of the same year. It was then referred to the Commerce

---

35 The privacy and electronic communications page can be viewed at http://www.ico.gov.uk/what_we_cover/privacy_and_electronic_communications.aspx for more and updated information about the spam regulations in the United Kingdom.

36 http://www.dcita.gov.au/communications_and_technology/policy_and_legislation/spam is government web site that contains full text of the Spam Act as implemented in Australia.

37 http://www.parliament.nz/en-NZ/PB/Legislation/Bills/1/d/f/1df7f2d2edc64dffba3500feec167939.htm is New Zealand Government web site that can be used to view full text.

Select Committee who reported back at the end of August 2006 with suggested amendments and enhancements. This Bill aims to prohibit the sending of unsolicited electronic messages (in the form of emails, text messages, or instant messages) of a marketing nature and provide a legislative basis to combat the growth of spam.

*Canada* does not yet have a dedicated anti-spam law, but the Personal Information Protection and Electronic Documents Act (PIPEDA)[38] covers online privacy in detail and contains many provisions relevant to email marketing. Although a government-initiated task force recommended specific anti-spam legislation in a 2005 report, this has yet to bear fruit in the form of an actual law. In order to build consumer trust and confidence in conducting e-business in Canada the Government of Canada is committed to establishing clear rules to protect the privacy of personal information in the new 'virtual' marketplace. This is being done through the implementation of Federal privacy legislation, and through development of a national policy on unsolicited consumer email, otherwise know as SPAM. The PIPEDA review has began on 20$^{th}$ November 2006.

In *India* CAUCE[39], The Coalition Against Unsolicited Commercial Email is an ad hoc, all volunteer organization, created by Netizens to advocate for a legislative solution to the problem of UCE. The Indian chapter of CAUCE is dedicated to nipping the spam problem in the bud in India, before it snowballs into a crisis.

US State laws against telemarketers were ineffective against out-of-state calls, and the US nationwide Do-Not-Call list has been ineffective against overseas calls. The many laws of the world's nations can be applied to their respective citizens, but not to a global Internet. Thus, the long arm of the law struggles to reach into cyberspace. Normal prosecutions require physical evidence, an accused, and a plaintiff. Yet spam can begin and end in cyberspace, email sources are easily spoofed, and, for spam, potential plaintiffs include everyone with an email address. Even if detected, a spam source can just reinvent itself under another name. Traditional law seems too physically restricted, too slow, and too impotent to deal with a dynamic, global information society. Further, Laws also tend to be too broad based.

---

38 The privacy in the Digital economy page on http://e-com.ic.gc.ca/epic/site/ecic-ceac.nsf/en/h_gv00045e.html provides full information on PIPEDA.

39 The web address of CAUCE is http://www.india.cauce.org/.

## Conclusion

The paper presented here discusses spam, its variants, potential problem created by spam and the current solutions for its control. The key approaches discussed in the paper to control spam are: i) technical, ii) economical, and iii) the legal. Owing to its complex character no single approach or a technique within that approach can be used to completely control spam from being transmitted. The complexity of the character of the spam is gauged by the fact that spam means different thing to different people and its dynamic nature. The spam control techniques applied will eventually be therefore user specific and needs to have adaptability. The paper explored various possibilities that can be employed to control the spam effectively. Whereas technical solutions will enforce spam control and will protect email users and ISP's to unsolicited mails, the economic and legal solutions may restrict spammers from over spamming. A uniform, yet user specific spam control technological and legal approach can prove to be a lasting spam control.